# Superconductivity of Bi Confined in an Opal Host

R. C. Johnson[1], M. D. Nieskoski[2†], S. M. Disseler[1], T.E. Huber[3] and M. J. Graf [1*]

[1] Department of Physics, Boston College, Chestnut Hill, MA 02467

[2] Department of Physics, Fairfield University, Fairfield, CT 06824

[3] Howard University, Washington, DC 20059

**ABSTRACT**

Superconductivity is observed in a composite of rhombohedral crystalline bismuth nanoparticles imbedded in an insulating porous opal host via electrical transport and AC magnetic susceptibility. The onset of superconductivity in this system occurs in two steps, with upper critical temperature $T_{c,U}$ = 4.1 K and lower transition temperature of $T_{c,L}$ = 0.7 K, which we attribute to the granular nature of the composite. The transition at $T_{c,U}$ is observed to split into two transitions with the application of a magnetic field, and has upper critical field extrapolated to $T$ = 0 K of $H_{c2,1}(0)$ = 0.7 T and $H_{c2,2}(0)$ = 1.0 T, corresponding to coherence lengths of $\xi_1(0)$ = 21 nm and $\xi_2(0)$ = 18 nm, respectively. We suggest that because of the lack of bulk-like states in the Bi nanoparticles due to confinement effects, superconductivity originates from surface states arising from Rashba spin-orbit scattering at the interface. This prospect suggests that nanostructured Bi may be an interesting system to search for Majorana fermions.

## I. INTRODUCTION

Bismuth is a semi-metal that exhibits a range of remarkable properties, including low number densities for the electron and hole carriers, extremely low effective masses, and very high mobilities.[1] Reducing the size of the bismuth sample causes a bulk band offset, with a semi-metal to semiconductor (SMSC) transition predicted to occur for sample dimensions less than the critical size $d_c \approx 50$ nm, with the exact size depending on the crystalline orientation relative to the confinement direction.[2] Moreover, spin-orbit effects are sufficiently strong at the surfaces of Bi to produce a new surface conduction band[3] that can dominate the transport properties for dimensions approaching $d_c$.

While bulk bismuth does not superconduct to temperatures of the order of 50 mK, it has been reported to superconduct with appropriate changes in the material morphology. Upon application of hydrostatic pressure a sequence of structural changes are observed to occur at pressures above 2.5 GPa, producing several distinct superconducting phases with transition temperatures of 3.9 K (2.5 GPa), 7.0 K (2.7 GPa), and 8.3 K (8 GPa).[4] Amorphous bismuth (a-Bi) films have been observed to superconduct at temperatures ranging from 0.5 K to 6 K,[5] depending on the film thickness and growth conditions. Studies of rhombohedral nanograins imbedded in a variety of solid matrices (e.g,, xenon and germanium) show that superconducting transition temperature $T_c$ is apparently dependent on both the grain size $L$ (with $T_c$ increasing linearly with decreasing $L$) and the host matrix.[6] The authors of Ref. 6 attribute the onset of superconductivity as a surface effect, presumably due to a modification of the carrier density of states at the grain/matrix boundary.

More recently, Bi granular nanowire arrays[7,8] and single crystal nanowires[9] have also been observed to become superconducting. These systems are intriguing because it is not clear what variation in the material parameters allows the nanowires to become superconducting. The observation of superconductivity in Bi nanowires is also surprising because a large number of studies on both single nanowires and arrays of nanowires have not shown evidence for superconductivity. Indeed, in Ref. 7 roughly half the samples studied exhibited superconductivity, with the others exhibiting 'local' superconductivity.[8] The fragility of superconductivity in nanowires seems to indicate that the phenomenon is associated with the surface. Possible explanations include variation in the carrier density as the system evolves from a semimetal to a semiconductor with transport properties dominated by conducting surface states,[3] an enhanced density of states due to the lower dimensionality,[6] or changes in the composition, crystallinity, or crystal structure at the surface. Superconductivity in crystalline Bi and its alloys may be of particular interest due to the possibility of observing Majorana fermions, which are considered strong candidates for storing quantum information and performing fault tolerant quantum computation, e.g., at a superconductor-topological insulator (TI) junction. Three dimensional topological insulators have surface states that are characterized by the existence of a gapless surface state that emerges because of the non-trivial $Z_2$ topology of the insulating bulk state and is protected against backscattering by time-reversal symmetry.[11] In the search for the Majorana particles several true TIs including $Bi_2Se_3$ (Refs. 12-15) were investigated and found to exhibit superconductivity but there is as yet no evidence for the Majorana particle. While Bi is not a true TI because it has a trivial topology, Hosur *et al*[16] examined a variety of systems, including PbTe, that are trivial TIs and found a criterion that does not require a true TI to support Majorana fermions.

With these long-term goals in mind, in order to further probe the origin of superconductivity in Bi nanowires it is of interest to investigate a system with a morphology intermediate between that of granular films and of nanowires. In this work we report electrical transport and AC magnetic susceptibility, supported by scanning electron microscopy (SEM), energy dispersive spectroscopy (EDAX), and x-ray diffraction (XRD) material characterization, of rhombohedral bismuth nanoparticles imbedded in an insulating porous opal host matrix formed from 160 nm $SiO_2$ spheres. We find the onset of superconductivity at a temperature of $T_{c,U}$ = 4.1 K. The transition occurs in two steps, with a lower transition temperature of $T_{c,L}$ = 0.7 K. We also present upper critical field data, and show that the coherence length is smaller than the typical particle size, consistent with the expected results for a granular superconductor. Further, we explore the possibility that superconductivity is associated with intrinsic surface charge carriers.

## II. EXPERIMENTAL

Porous silica opals were used as the host matrix, and the fabrication of composite samples by pressure injection of molten Bi has been described in detail elsewhere.[17,18] X-ray diffraction measurements were performed, and the sharpness and angular position of the diffraction lines that are observed show that the nanoparticles have a rhombohedral crystal structure with lattice parameters similar to those of bulk Bi but the distribution of lines indicate that there are no preferred reflections, from which we infer that the nanoparticles are crystalline with no preferential orientation. We have also carried out a compositional analysis via EDAX on our scanning electron microscope and we only resolve Bi, Si, and O components, with atomic percentages of 9 %, 22 %, and 68 %, respectively. There appears to be an excess of oxygen, which we believe is due to adsorption on the surface during the processing of the sample for the SEM studies. Assuming the excess oxygen is due to surface adsorption, the Bi mass fraction is calculated to be 53 %, corresponding to a volume fraction of 24 %. This value is consistent with estimates from direct measurement of the sample density, which yield a Bi volume fraction of 24

%. We conclude that the Bi nearly completely fills the voids in the opal host, as both these estimates are close to the theoretical limit of 26 % open volume in an ideal FCC lattice of close-packed spheres. For an FCC opal lattice the metal imbedded within the host can be modeled as spheres occupying the tetrahedral and octahedral sites, with diameters of $0.414D$ and $0.225D$, respectively, and with cylinders of diameter $0.155D$ linking the spheres to form a network.[19] Thus the three characteristic diameters for the Bi nanoparticles in our sample are 66 nm, 36 nm, and 25 nm; note that these sizes are near or below the expected size to induce the SMSC transition. Using the simple model presented in Ref. 19, where the Bi network consists of 66 nm (8 per unit cell) and 36 nm diameter spheres (4 per unit cell) connected by 25 nm cylinders (32 per unit cell), we estimate the Bi volume fractions to be 60 %, 5%, and 35 %, respectively.

For electrical transport measurements, six wires were attached to the sample using silver epoxy, and four-terminal electrical transport measurements were taken using an LR-700 AC resistance bridge. Varying the lead configuration for the four-terminal measurement showed that the resistance scaled linearly with distance between voltage contacts, confirming that our measurements are not dominated by cracks or other large scale imperfections in the sample. The samples were cooled to low temperatures using either a gas flow cryostat (1.6 K $< T <$ 300 K) or a $^3$He adsorption refrigerator (0.3 K $< T <$ 5 K). Current levels were varied to ensure that no sample heating occurred. Both refrigerators were used in conjunction with a 9 T superconducting magnet.

Measurements of the AC susceptibility above 2 K were carried out in an Oxford Instruments MagLab Magnetic Measurements system, using the two-coil mode to minimize the background contribution to the measured signal. The system was calibrated with a superconducting Nb sphere. A homebuilt susceptometer was used at temperatures below 3 K in our [3]He adsorption refrigerator, and was calibrated with a superconducting sample and cross-calibrated with data from the MagLab system. Measurements were taken at a frequency of 500 Hz and drive excitations of 2 G and 0.1 G in the MagLab and homebuilt susceptometers, respectively. Variation of the frequency and excitation showed that no sample heating occurred. Several pieces from the same sample were studied, with masses in the range 30 − 50 mg; all yielded similar results. The typical demagnetization factor based on the bulk sample dimensions was approximately 0.6.

## III. RESULTS

In Fig. 1 we show the resistivity as a function of temperature for one of our samples; similar data were obtained for other pieces. The resistivity at room temperature is 36 μΩ-cm compared to the typical 1 μΩ-cm for bulk bismuth, and consistent with the roughly order of magnitude increase expected due mean free path limitations,[20] surface states,[3] and quantum size effects.[2] The resistivity increases with decreasing temperature down to approximately 30 K, below which it is nearly constant with a further decrease in temperature. From the negative $dR/dT$ we infer that the resistivity of the Bi confined within the opal host is dominated by the small Bi nanoparticles, which have sizes comparable to the critical size required to induce the SMSC transition. Indeed, the temperature dependent resistance is nearly identical to that reported for Bi nanowire arrays with diameters below 50 nm, that is, below the SMSC transition.[21] A dramatic and abrupt decrease in the resistivity is seen below 4.3 K, from which we extract a transition temperature $T_{c,U}$ = 4.3 K, although it does not approach zero down to 2 K. A second drop of the resistivity begins below 2 K, with a lower transition temperature at $T_{c,L} \approx 0.7$ K. Both transition temperatures were determined from a peak in the temperature derivative of the resistance. The resistivity is nearly zero by our base temperature of $T$ = 0.3 K. A test of the sensitivity to drive current showed that the transition width at $T_{c,U}$ is somewhat sensitive to excitation level (inset, Fig. 1).

In Fig. 2a we show the low temperature variation of the scaled resistivity with temperature in several magnetic fields; the data has been normalized to the zero-field resistance value at $T$ = 4.5 K. The lower transition ($T_{c,L}$) shifts rapidly with field to below our minimum temperature of 0.3 K. The upper transition ($T_{c,U}$) is observed to broaden with applied field. Closer inspection of the derivative of the curve in the transition region shows two peaks, both associated with the upper temperature transition (see Fig. 2b). These two peaks broaden considerably with applied field, and we cannot follow the field dependence of the transition temperatures below about 2 K ($H \sim 0.5$ T). However, a rough plot of the critical field versus temperature yields two curves which can be reasonably fit to the mean-field form for the temperature dependent upper critical field $H_{c2}(T)$, as shown in Fig. 3. These fits yield zero temperature critical field values of $H_{c2,1}(0)$ = 0.7 T and $H_{c2,2}(0)$ = 1.0 T, corresponding to coherence lengths of $\xi_1(0)$ = 21 nm and $\xi_2(0)$ = 18 nm, respectively. These values are comparable to the particle size, as expected for a granular superconductor, and will be discussed further in the next section.

Results for the zero-field AC susceptibility for a sample below 6 K are shown in Figure 4. An abrupt change is observed at 4.3 K, which, taken with the resistivity data presented above, we associate with the onset of superconductivity. This change was observed on different pieces of sample in both the MagLab and homebuilt susceptometers. Calculating the superconducting volume fraction relative to the estimated total Bi volume, $f_s$, at 2 K we find the small value $f_s$ = 1.5(5) %. The relatively large error is due to the small sample size and the error involved in determining the volume fraction of Bi in the composite sample. Below 2 K, the susceptibility begins to decrease once again. At our lowest temperature $T = 0.5$ K, below $T_{c,L}$, the susceptibility is still decreasing, and $f_s$ is calculated to be 8.0(5) % at this temperature.

Finally, we note that limited measurements of the low field modulated microwave absorption[22] (LFMA) have recently been made on a piece from the same samples studied here.[23] These measurements reveal the onset between $T = 4.5$ K and 3.9 K of a non-resonant absorption signal centered at zero field. The abrupt onset of such a signal is typically associated with changes in surface conductivity, fluxon dynamics, or tunneling currents between grains in superconductors. More detailed studies will be helpful in determining the role of granularity in the observed superconducting properties and the details of the temperature dependent penetration depth in this system.

## IV. DISCUSSION

Typically, a two-step resistive transition as shown in the Fig. 1 is associated with granular superconductivity. In this case one first achieves local superconductivity in individual small grains, but fluctuations prevent phase coherence between the Josephson-coupled grains until a lower temperature.[24] Such a large discrepancy between local and global transition temperatures implies very weak coupling between superconducting grains. The two-step transition in the susceptibility is unusual, however, since macroscopic screening currents cannot flow on the sample surface when there is no global superconductivity. A similar two-step transition in the susceptibility was observed for gallium imbedded in an opal host (sphere diameter of 250 nm).[25] In this case the two transitions were attributed to a structural modification that occurs in confined gallium at temperatures below 150 K, resulting in mesoscopically inhomogeneous sample with regions of different crystalline structure, and, therefore, different superconducting transition temperatures. The many experiments on Bi nanowires and nanoparticles over the past decade have not shown evidence for such a structural modification,[26] so this scenario seems unlikely to us. More likely is the possibility that the individual grains in fact produce a superconducting diamagnetic response at $T_{c,U}$, as calculated in Refs. 27 and 28, and recently observed in coated lead nanoparticles[29]. Because the nanoparticle dimensions $R$ are comparable to, or smaller than, the London penetration depth $\lambda_L$, a nanoparticle's susceptibility is reduced relative to the bulk superconducting diamagnetic susceptibility according to[27]

$$\frac{\chi_{nano}}{\chi_{bulk}} = \frac{3}{20}\frac{R^2}{\lambda_0^2} \qquad , \qquad \text{Eq. 1}$$

where we assume that $R$ is comparable to or larger than the zero-temperature bulk value of the coherence length $\xi_0$; this assumption and the assumption that $R << \lambda_L$ will be justified later in this discussion. For typical type-I superconductors and $R = 66$ nm this reduction is in the range $0.1 - 1\%$. A more exact analysis is not possible since bulk Bi is not superconducting, although we present a calculation based on surface carriers below. This change is comparable to our observed change of -1.5(5) % at $T_{c,U}$ and despite the relatively large uncertainty in the estimated and measured quantities, we can infer that a large fraction, and possibly all, of the Bi becomes superconducting at $T_{c,U} = 4.1$ K. The large change in susceptibility below 1 K indicates the onset of phase coherence between the weakly coupled nanoparticles, and therefore macroscopic shielding currents that restore the susceptibility to the anticipated bulk superconducting value.

This picture is of course consistent with our observations of the two-step resistive transition in zero field. At present we do not have an explanation for the origin of the two transitions observed in small fields at $T_{c,U}$, although possible effects include a size dependence of the transition temperature or slightly different demagnetization factors for the spherical and cylindrical nanoparticles (approximate aspect ratio of 3:1).

The main question remains: Why are crystalline Bi nanoparticles superconducting? We critically outline some possible mechanisms by which superconductivity can occur in the nanoparticles even though bulk Bi is not superconducting. First, it is possible that the crystal structure changes from the room temperature rhombohedral structure observed via x-ray diffraction. Causes might include bulk changes induced by enhanced pressure due to the particle curvature, or by surface reconstruction effects. The former mechanism is unlikely, as any reasonable estimate for the curvature induced pressure via the Rayleigh formula yields pressures an order of magnitude lower than the 2.5 GPa needed to induce a phase transition. Moreover, the observed transition temperature $T_{c,U}$ = 4.1 K is significantly lower than the expected transition temperatures 7 – 8 K for the stable high-pressure phases of Bi.

It is also possible that the surface regions of the nanoparticles are amorphous, e.g., with a high density of defects, and so become superconducting in analogy with amorphous thin films. Studies on quench condensed a-Bi thin films[30-33] show that they become superconducting at film thicknesses greater than approximately 1 - 2 nm for homogeneous a-Bi. If we assume that our nanoparticle surfaces were amorphous to a depth of roughly 2 nm, it would account for approximately 12% of the total Bi in the network. XRD measurements on the sample show that the Bi in our samples is crystalline and there is no indication of a-Bi, although it is not clear that such a thin layer would show up in XRD. However, a-Bi is characterized by a much lower resistivity ρ than for crystalline Bi and has a positive dρ/dT, so that a 2 nm thick surface of a-Bi would dominate the transport over crystalline semiconducting Bi "cores". In fact we find a large resistivity and a negative dρ/dT, leading us to believe that the observed transport properties are due to crystalline semiconducting Bi nanoparticles with sizes below the critical value $d_c$ for the SMSC.

Finally, it may be that the superconductivity results from an enhanced density of states at the crystal surface, as put forth in Refs. 6 and 7. This explanation is appealing, as transport[3] and ARPES[34] measurements have shown the existence of intrinsic surface state carriers, which are predominant for nanoparticle dimensions less than about 100 nm.[21] These states exist in region of about 20 nm within the surface – essentially all of the volume for nanoparticles studied in this work. The properties of the surface state carriers were estimated from transport and thermopower measurements in Ref. 21: the effective mass $m* = 0.2m_e$, the carrier density $n = 1.3 \times 10^{19}$ m$^{-3}$, and the Fermi velocity $v_F = 5 \times 10^4$ m/s. Using the standard expressions for the BCS coherence length $\xi_0$ and London penetration depth $\lambda_L$ (see for example Ref. 27)

$$\xi_0 = 0.18 \frac{\hbar v_F}{k_B T_c} \quad , \quad \lambda_L = \left( \frac{m*}{\mu_0 \, n \, e^2} \right)^{\frac{1}{2}} \qquad \text{Eq. 2}$$

we estimate that for the surface carriers, $\xi_0 \approx 17$ nm and $\lambda_L \approx 210$ nm. Note that $\xi_0 \sim R \ll \lambda_L$. Revisiting the zero-temperature coherence lengths extracted from the upper critical field data, $\xi_1(0) = 21$ nm and $\xi_2(0) = 18$ nm, we can now evaluate the expression for a 'dirty' type-II superconductor, $\xi(0) = (\xi_0 l)^{1/2}$, where $l$ is the electronic mean free path. We find $l_1 \approx 26$ nm and $l_2 \approx 19$ nm, consistent with $l$ being limited by scattering at boundaries of the grains studied here, with dimensions in the range 25 – 66 nm. The estimated reduction in the diamagnetic susceptibility from Eq. 1 is found to be 0.3%, in fair agreement with the observed change in susceptibility at $T_{c,U}$ given the large uncertainties involved. We note that if indeed the surface states are superconducting, one might expect some unusual properties associated with them due to the presence of strong Rashba spin-orbit scattering,[35] and as described above in the context of observing Majorana fermions.

Still, for the majority of Bi nanowires studied in earlier works, these same surface states have not been observed to become superconducting. If these states were in fact responsible for superconductivity in these Bi nanocrystals, then the superconducting state would need to be very sensitive to slight changes in the surface environment in order to account for the rather sporadic occurrence of superconductivity in nanowires and nanograins.

**V. SUMMARY**

In conclusion we have observed superconductivity in rhombohedral bismuth nanoparticles imbedded in an insulating porous opal host matrix via resistivity and AC magnetic susceptibility measurements. While the origin of superconductivity in Bi remains uncertain, we point out the possibility that intrinsic surface states are responsible for superconductivity. Further detailed studies on the reproducibility and effects of varying the particle size will be helpful in clarifying the origins of superconductivity in Bi-opal composites, and in other bismuth nanoparticle and nanowire systems. It would also be of interest to search for superconductivity in nanostructured Sb, which is a true topological insulator and for which confinement can also eliminate the bulk carriers, leaving only surface carriers.

**ACKNOWLEDGEMENTS**

We thank Prof. A. Zakhidov for generously providing the sample studied, and Mr. A. Howard for reporting the preliminary results of his low field microwave absorption measurements on this sample to us. We also acknowledge technical assistance provided by Mr.

Tom Hogan. This work was supported in part by National Science Foundation grant REU DMR-0649169 (MDN). T.E.H acknowledges support the National Science Foundation under Grant No. NSF-DMR-0839955 and by the U.S. Army Research Office Materials Science Division under Grant No. W911NF-09-1-05-29.

[†] Present address, Thayer School of Engineering at Dartmouth, Hanover, NH 03755

* grafm@bc.edu

**FIGURE CAPTIONS**

Figure 1. Resistance vs temperature for Bi-opal composite samples, scaled to the room temperature resistance value $R_{300K}$. Inset: sensitivity of the resistive transition near 4 K to excitation current.

Figure 2. (a) Scaled resistance versus temperature in several fixed magnetic fields; the curves are not offset, but reflect the positive magnetoresistance of the sample. (b) The derivative of the curves in (a) with respect to temperature; curves are offset for clarity.

Figure 3. Upper critical field diagram for the split two transitions observed in field for $T_{c,U}$. Solid lines are fits to the form $H(T) = H_0 [1-(T/T_{c,U})^2]$

Figure 4. In-phase portion of the AC Volume Susceptibility vs. temperature for the Bismuth in the Bi-opal composite samples.

**Fig. 1**

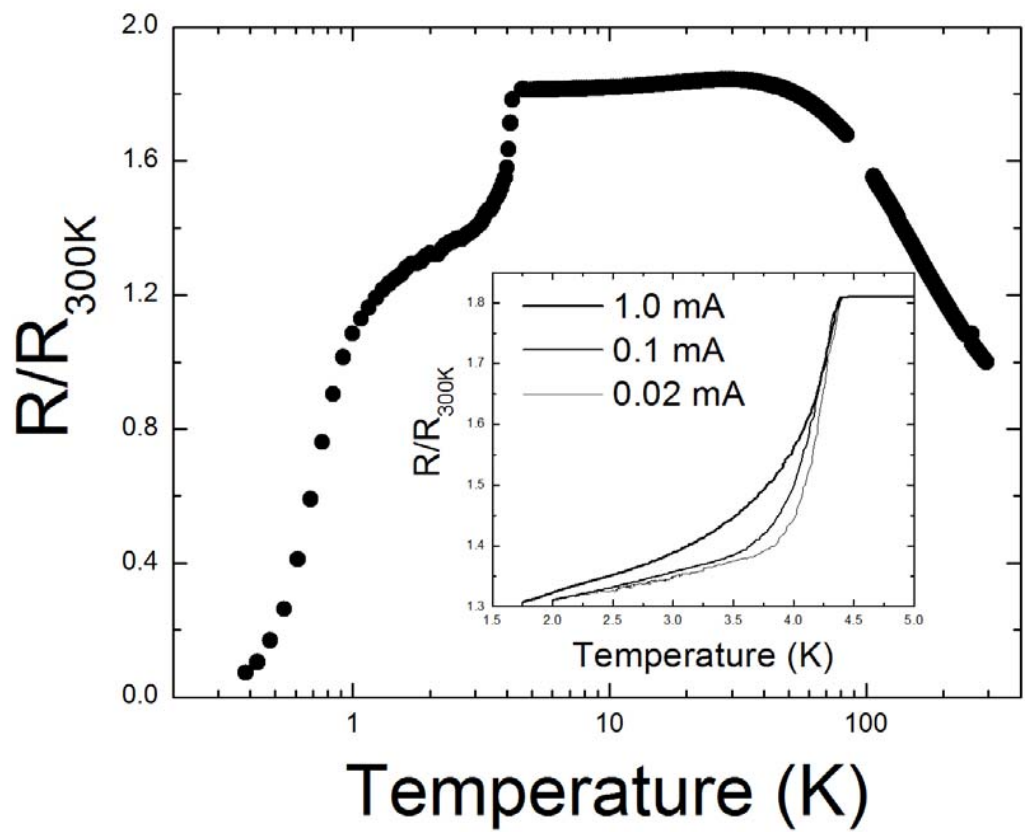

**Fig. 2**

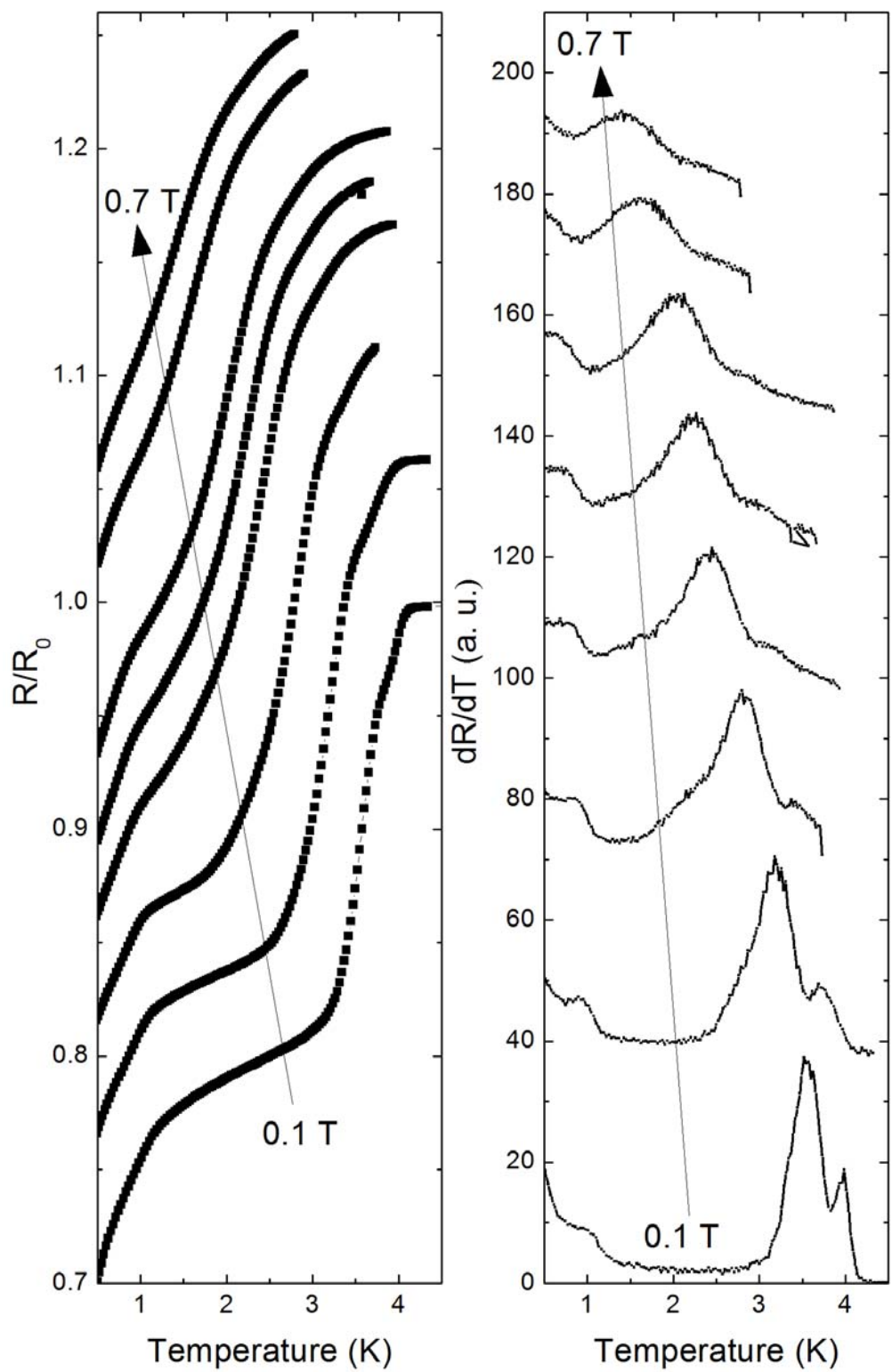

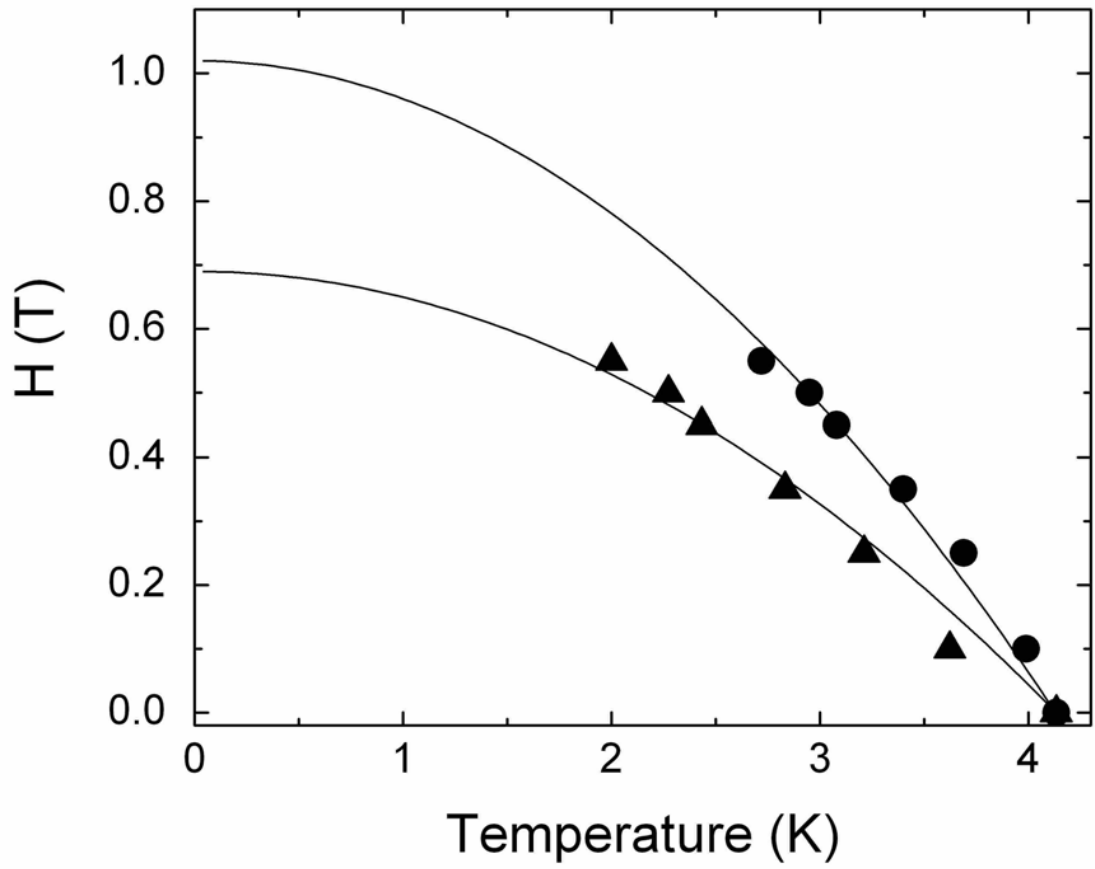

**Fig. 4**

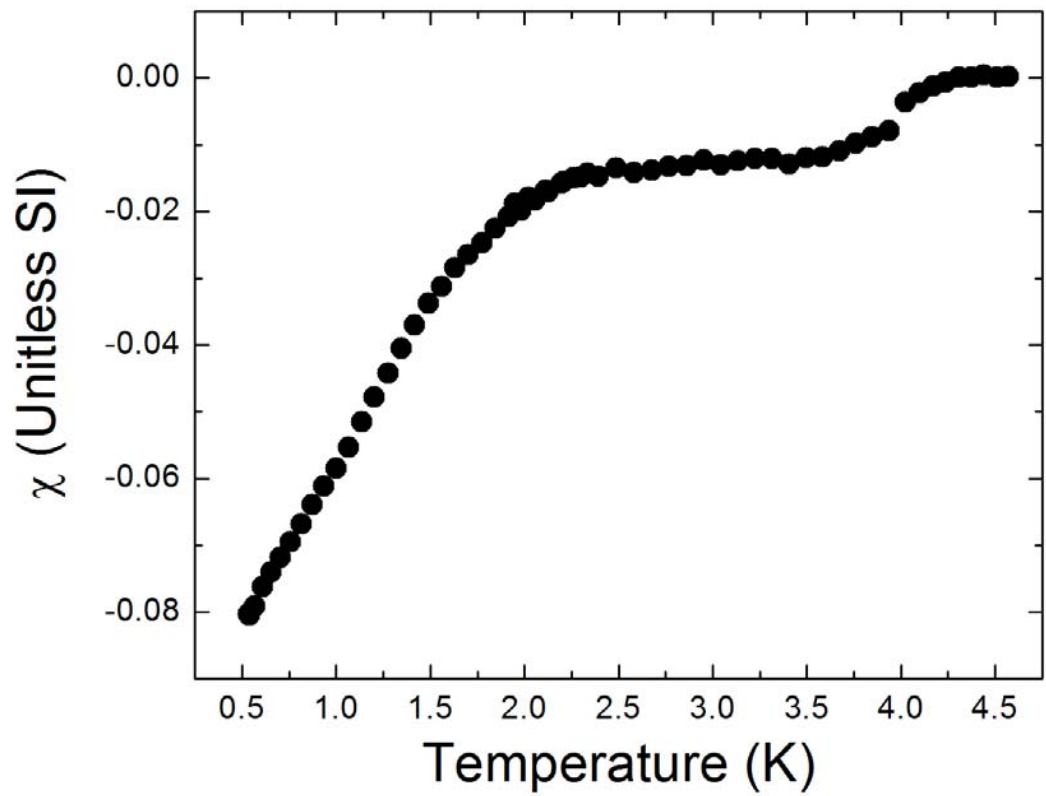